\documentclass[format=sigconf]{acmart}

\usepackage{booktabs} % For formal tables
\setcopyright{rightsretained}

\copyrightyear{2018}
\acmYear{2018}
\setcopyright{acmlicensed}
\acmConference[MSR '18]{MSR '18: 15th International Conference on Mining Software Repositories }{May 28--29, 2018}{Gothenburg, Sweden}
\acmBooktitle{MSR '18: MSR '18: 15th International Conference on Mining Software Repositories , May 28--29, 2018, Gothenburg, Sweden}
\acmPrice{15.00}
\acmDOI{10.1145/3196398.3196447}
\acmISBN{978-1-4503-5716-6/18/05}
\usepackage[utf8]{inputenc}
\usepackage{url}
\usepackage{enumitem}
\usepackage{flushend}
\usepackage{graphicx}
\usepackage{amssymb}
\usepackage{subcaption}
\usepackage{xspace}
\usepackage{hyperref}
\newcommand{\etal}{et al.\xspace}

\newcommand{\fig}[1]{Fig.~\ref{#1}}
\newcommand{\tab}[1]{Table~\ref{#1}}

\begin{document}
%\title{Towards Identifying Paid Open Source Developers -\\
%  A Case Study with Mozilla Developers}

\title{Towards Automatically Identifying Paid Open Source Developers}


\author{Maëlick Claes}
\affiliation{M3S, ITEE, University of Oulu, Finland}
\email{maelick.claes@oulu.fi}

\author{Mika Mäntylä}
\affiliation{M3S, ITEE, University of Oulu, Finland}
\email{mika.mantyla@oulu.fi}

\author{Miikka Kuutila}
\affiliation{M3S, ITEE, University of Oulu, Finland}
\email{miikka.kuutila@oulu.fi}

\author{Umar Farooq}
\affiliation{M3S, ITEE, University of Oulu, Finland}
\email{umar.farooq@oulu.fi}

\begin{abstract}
%!TEX root = paper.tex

Open source development contains contributions from both hired and
volunteer software developers. Identification of this status is
important when we consider the transferability of research results to
the closed source software industry, as they include no volunteer
developers. While many studies have taken the employment status of
developers into account, this information is often gathered manually
due to the lack of accurate automatic methods. In this paper, we
present an initial step towards predicting paid and unpaid open source
development using machine learning and compare our results with
automatic techniques used in prior work. By relying on code source
repository meta-data from Mozilla, and manually collected employment
status, we built a dataset of the most active developers, both
volunteer and hired by Mozilla. We define a set of metrics based on
developers' usual commit time pattern and use different classification
methods (logistic regression, classification tree, and random
forest). The results show that our proposed method identify paid and
unpaid commits with an AUC of 0.75 using random forest, which is
higher than the AUC of 0.64 obtained with the best of the previously
used automatic methods.

\end{abstract}

\maketitle

%!TEX root = paper.tex

\section{Introduction}

As open source software simplifies the acquisition of large amounts of
data related to software development, it has been the subject of
numerous repository mining studies. Despite the large focus on open
source software in the research community, and long existing studies
showing that large companies invest money in open source
development~\cite{Ghosh2002, Lakhani2005}, many empirical studies
still often assume that the software industry is split between open
source projects developed mainly by non-paid hobbyist
developers~\cite{Torvalds2001, Madey2002, KRISHNAMURTHY2014632,
  Choi2013, wu2013, AlMarzouq2015, JEMS12053,
  Jergensen2011-onionpatch}, and closed-source projects developed by
hired developers paid by companies.

Because of the differences between volunteer work and paid work, a
common threat to validity in repository mining studies relying on open
source data, is that findings may not hold for traditional
closed-source projects~\cite{Godfrey2000} and vice versa. This is
particularly true when the research focuses on social, human or
behavioral aspects of software development. Another related issue is
that some findings about open source software may only hold when the
developers are volunteer. For example,
Herraiz~\etal~\cite{Herraiz2006} found that the onion model doesn't
apply to core-developers working for companies but only for
volunteers. There are also fundamental differences between paid and
volunteer work as volunteer work tend to be performed by individual
with greater well-being and increase well-being at the same
time~\cite{thoits2001} and that a motive for volunteer work is 
different from hired work ~\cite{PROUTEAU2008314}.

Because of earlier results we obtained studying working hours of
Mozilla developers~\cite{claes2017abnormal, Claes2018ICSE}, we believe
there are some important differences between paid closed-source, paid
open source software and volunteer open source software
development. In order to increase the industrial relevance,
generalizability to correct population and credibility of mining
software repositories studies we provide a first step towards
(semi-)automatically identifying whether developers are paid or not.

To achieve this goal, we rely on code repository meta-data and
manually extracted information about developer employment status in
order to identify which characteristics better predict whether a
developer or a commit should be considered as ``hired'' or
``volunteer''. We focus on the Mozilla projects because they contain a
large amount of developers paid by the Mozilla Foundation. This
simplifies the process of manually gathering employment status of
developers needed to train and test our classifiers. We then compare
our results with the ones obtained by applying techniques used in the
literature we are aware of~\cite{claes2017abnormal, Riehle2014}.

%%% Local Variables:
%%% mode: latexmk
%%% TeX-master: "paper.tex"
%%% End:

%!TEX root = paper.tex

\section{Related Work}\label{sec:related}

The most reliable technique to identify paid open source developers is
gathering data manually by searching for personal information on web
search engines~\cite{Teixeira2015}. Indeed, it is common for open
source developers to advertise themselves and mention for which
company they work on websites such as LinkedIn, GitHub or their
personal websites. However, it is long and laborious and cannot scale
to thousands of developers. Unfortunately, the only alternative
methods we found in the existing literature are the use of the time of
activity~\cite{Riehle2014} and e-mail address domain
names~\cite{claes2017abnormal}.

Riehle \etal~\cite{Riehle2014} studied the amount of paid work in
various open source projects, including the Linux kernel. They
considered that any commit made on a weekday between 9am and 5pm was
paid work and work outside this time period was voluntary
work. However, this technique completely ignores unemployment,
flexible working hours and overwork. Open source developers can
potentially be students or unemployed and thus contribute outside the
what is considered as regular office hours. Additionally, developers
can also have flexible or irregular working hours. These two issues
can potentially hinder the precision and the recall of the
results. Moreover, even developers with regular office hours can be
subject to overwork. This can also be a non negligible source of false
negatives.

In our previous work~\cite{claes2017abnormal}, we studied abnormal
working hours in Mozilla Firefox. In order to easily identify paid
developers, we used the email address provided in the commit meta-data
and considered that every developer using an email registered at
mozilla.com is a Mozilla employee. However, the technique is not
perfect as we found Mozilla employees who do not use their Mozilla
e-mail address in commits. On the other hand, active volunteers could
also own and use a Mozilla email address.

%%% Local Variables:
%%% mode: latexmk
%%% TeX-master: "paper.tex"
%%% End:

%!TEX root = paper.tex

\section{Methodology}\label{sec:methodo}

\subsection{Data extraction}

To answer the research questions, we mined data from Mozilla's
Mercurial repositories\footnote{\url{https://hg.mozilla.org}}. We
extracted the history of all commits using
\emph{git-remote-hg}\footnote{\url{https://github.com/felipec/git-remote-hg}}
 and the \emph{GrimoireLab}
tools\footnote{\url{https://grimoirelab.github.io/}}.

In addition, we extracted issue comments from Mozilla's Bugzilla
repository\footnote{\url{http://bugzilla.mozilla.com}} (i.e., the
database containing reported issues, such as bug reports or feature
requests). In order to identify which commits from Mozilla's code
repositories are related to which Mozilla sub-project (e.g., Firefox
for Desktop vs. Firefox for Android), we linked commit messages to the
corresponding issue report by looking for an issue identifier in a
given message. Out of the 396,180 extracted commits, 330,078 were
successfully linked to a bug issue. After linking, we then filtered
the commits to only keep the ones related to the following major
\emph{products}: \emph{Firefox}, \emph{Core}, \emph{Firefox OS},
\emph{Firefox for Android}, \emph{Thunderbird} and \emph{SeaMonkey}.

In order to study the individual developers,
we performed a basic merging of the different authors' identities. We
first cleaned the name and email used in the version control system's
author field. Then we grouped together identities using the same name
or email addresses. Finally, two of the authors manually checked the
result in order to avoid any false positive.

In order to obtain ground truth about hired developers at Mozilla, we
ran a manual background check for the developers with more than 100
commits, leaving us with 391 developers (out of 2,755). 261 (66.8\%)
of those developers were hired by Mozilla and made 87\% out of all the
Firefox commits.

We also collected periods of employment, which we could retrieve for
212 of the hired developers, and found 16 developers who contributed
both as volunteers and hired developers. 6 of them have committed more
as a volunteer than as a hired and were considered as volunteers in
the data set. For the others, the amount of commits made as a
volunteer was relatively small (less than 20\%) and we considered them
as hired developers.

\subsection{Classification algorithms and metrics}

First, we use three algorithms to predict developer employment status. First
we use logistic regressions because our predicted variable is only
binary (hired or volunteer). Then, we use classification and
regression tree (CART) models using the \emph{rpart} R
package~\cite{rpart2015package}. Finally, we also use random
forests. While random forest usually gives better results than a
single decision tree, \emph{rpart} allows us to visualize the
decision tree and better understand which features better predict
employment status. We built all of these models using a repeated
10-fold cross validation with the \emph{caret} R
package~\cite{kuhn2008caret}.

Second, we define a set of metrics about developer commit activity
that will be used as features for our classification algorithms. These
metrics are summarized in \tab{tab:metrics}. For the metrics related
to the time of the day or time of the week, we took into account the
timezone given in the commit timestamp and thus consider the
developer's local time.

\begin{table}[!htpb]
  \centering
  \caption{List of metrics computed for each developer}
  \label{tab:metrics}
  \begin{tabular}{|l|l|}
    \hline
    Metric & Description \\
    \hline
    \emph{period} & n of days between first and last commit \\
    \emph{days} & n of days with at least one commit \\
    \emph{weeks} & n of weeks with at lest one commit \\
    \emph{timediff} & median of days between successive commits \\
    \emph{commits} & n of authored commits \\
    \emph{loc per commit} & median loc modified per commit \\
    \emph{weekend} & \% of commits during the weekend \\
    \emph{night} & \% of commits between midnight and 6am \\
    \emph{morning} & \% of commits between 6am and noon \\
    \emph{afternoon} & \% of commits between noon and 6pm \\
    \emph{evening} & \% of commits between 6pm and midnight \\
    \emph{office} & \% of commits between 8am and 5pm \\
    \emph{most active hour} & h of day with highest amount of commits \\
    \emph{beginning regular} & h of day when weekday activity starts \\
    \emph{length regular} & Length of weekday activity period \\
    \emph{end regular} & h of  day when weekday activity ends \\
    \hline
  \end{tabular}
\end{table}

%%% Local Variables:
%%% mode: latexmk
%%% TeX-master: "paper.tex"
%%% End:

\section{Empirical analysis}\label{sec:analysis}

In this section, we first build logistic regression, decision tree and
random forests models using a 10-fold cross validation. Secondly, we
compare these results with simpler techniques used in the
literature. Then, we use the information from the most active
developers to predict whether individual commits are paid. Finally, we
discuss the differences obtained between the different models built
and the different simple techniques used in the literature.

\subsection{Prediction of employment status of individual developers}

First we ran our three classification algorithms with a 10 repeated
10-fold cross validation using all the metrics defined in
\tab{tab:metrics} as features. \tab{tab:measures_all} reports the ROC
area under the curve, precision and recall of the three different
models.

\begin{table}[!htpb]
  \centering
  \caption{Predicting employment status using all metrics}
  \label{tab:measures_all}
  \begin{tabular}{l|l|l|l}
    Classifier & ROC AUC & Precision & Recall \\
    \hline
    \emph{logit} & 0.73 & 0.751 & 0.884 \\
    \emph{rpart} & 0.65 & 0.735 & 0.879 \\
    \emph{randomforest} & 0.77 & 0.767 & 0.859
  \end{tabular}
\end{table}

In addition, we also ran the same three algorithms using the subset of
features not related to the period or amount of activity of a
developer. We left out \emph{commits}, \emph{days}, \emph{weeks} and
\emph{period} from \tab{tab:metrics}. Because we want to be able
to train a model with a relatively small number of developers,
potentially the most active, and still be able to predict the outcome
for less active developers. The performance metrics, using a 10
repeated 10-fold cross validation, are reported in
\tab{tab:measures_subset}.

\begin{table}[!htpb]
  \centering
  \caption{Predicting author
    employment status without using metrics related to the number of
    commits or periods of activity.}
  \label{tab:measures_subset}
  \begin{tabular}{l|l|l|l}
    Classifier & ROC AUC & Precision & Recall \\
    \hline
    \emph{logit} & 0.68 & 0.732 & 0.897 \\
    \emph{rpart} & 0.63 & 0.73 & 0.873 \\
    \emph{randomforest} & 0.75 & 0.756 & 0.881
  \end{tabular}
\end{table}

Comparing \tab{tab:measures_all} and \tab{tab:measures_subset}, we
observe that the performances are quite similar for \emph{rpart} and
\emph{randomforest}. This means that the importance of the number of
commits, days, weeks and period of activity is not critical to predict
the employment status of a developer. It also means it should be
possible to train a model with only a small number of developers and
still be able to predict the employment status of the other developers

\subsection{Comparison with simpler automatic techniques}

In order to assess the performance of the three models, we computed
the same performance metrics obtained with simple automatic techniques
used in the existing literature to detect paid developers. These are
reported in \tab{tab:measures_basic}.

\begin{table}[!htpb]
  \caption{Predicting employment status without machine learning}
  \label{tab:measures_basic}
  \centering
  \begin{tabular}{l|l|l|l}
    Classifier & ROC AUC & Precision & Recall \\
    \hline
    \emph{allhired} & 0.5 & 0.668 & 1 \\
    \emph{email} & 0.64 & 0.772 & 0.701 \\
    \emph{95\%officehours} & 0.5 & NA & 0 \\
    \emph{5\%officehours} & 0.5 & 0.668 & 1 \\
    \emph{50\%officehours} & 0.63 & 0.75 & 0.805
  \end{tabular}
\end{table}

First, because only one third of all the considered authors are
volunteers, we computed the performance metrics when considering all
authors as hired (\emph{allhired}). Then, we relied on the domain name
of the email address used by developers. Like in our previous study
about developer working hours in Firefox~\cite{claes2017abnormal}, we
considered as paid, the developers who made at least one commmit with
a mozilla.com email address (\emph{email}).

Finally, we also tried to consider the approach used by Riehle
\etal~\cite{Riehle2014}. They considered that hired developers are
developers for which at least 95\% of their commits were made during
regular office hours (9am-5pm) and volunteers less than 5\% of their
commits. All developers having made more than 5\% but less than 95\%
during the same time interval were considered as having a mixed
status.

Because we only consider a developer as hired or volunteer, we
computed the performance metrics for the case where hired developers
are those with more than 95\% of their commits during office hours
(\emph{95\%officehours}), and the case case where hired developers are
those with more than (\emph{5\%officehours}). These are equivalent as
considering either all developers as volunteers or as hired. Indeed
all considered developers had between 5\% and and 95\% of their
commits made during regular office hours. In addition we also consider
the case where a hired developer is a developer with more than 50\% of
commits during office hours (\emph{50\%officehours}).

Overall, all of these simple automatic techniques give performances
below the random forest classifier in terms of AUC.

\subsection{Predicting paid commits}

To detect individual paid commits instead of individual paid
developers, we build our \emph{logit}, \emph{rpart} and
\emph{randomforest} models using the information from the 67 most
active developers who have contributed, altogether, at least 50\% of
all of the considered commits. We tested the performance of these
models on all the commits and on the
commits of the least active developers
(\tab{tab:measures_commits}). We also computed the performance metrics
when considering as paid commits, all the commits (\emph{allpaid}),
the commits made by a developer with at least a Mozilla email address
(\emph{email}) as done in our previous study~\cite{claes2017abnormal},
and the commits made during regular office hours (\emph{officehours})
as done by Riehle \etal~\cite{Riehle2014}.

\begin{table}[!htpb]
  \caption{Models built with the most active developers and tested on all commits (top) and on commits of the least active developers (bottom).}
  \label{tab:measures_commits}
  \centering
  \begin{tabular}{l|l|l|l}
    Classifier & ROC AUC & Precision & Recall \\
    \hline
    \emph{logit} (all) & 0.683 & 0.86 & 0.865 \\
    \emph{rpart} (all) & 0.766 & 0.843 & 0.954 \\
    \emph{randomforest} (all) & 0.87 & 0.879 & 0.957 \\
    \emph{allpaid} (all) & 0.5 & 0.781 & 1 \\
    \emph{email} (all) & 0.594 & 0.853 & 0.672 \\
    \emph{officehours} (all) & 0.467 & 0.743 & 0.414 \\
    \hline
    \emph{logit} (least active) & 0.608 & 0.781 & 0.714 \\
    \emph{rpart} (least active) & 0.724 & 0.768 & 0.935 \\
    \emph{randomforest} (least active) & 0.69 & 0.754 & 0.904 \\
    \emph{allpaid} (least active) & 0.5 & 0.704 & 1 \\
    \emph{email} (least active) & 0.639 & 0.813 & 0.702 \\
    \emph{officehours} (least active) & 0.448 & 0.644 & 0.393
  \end{tabular}
\end{table}

\subsection{Discussion}

\begin{figure}[!htpb]
  \centering
  \includegraphics[width=\columnwidth]{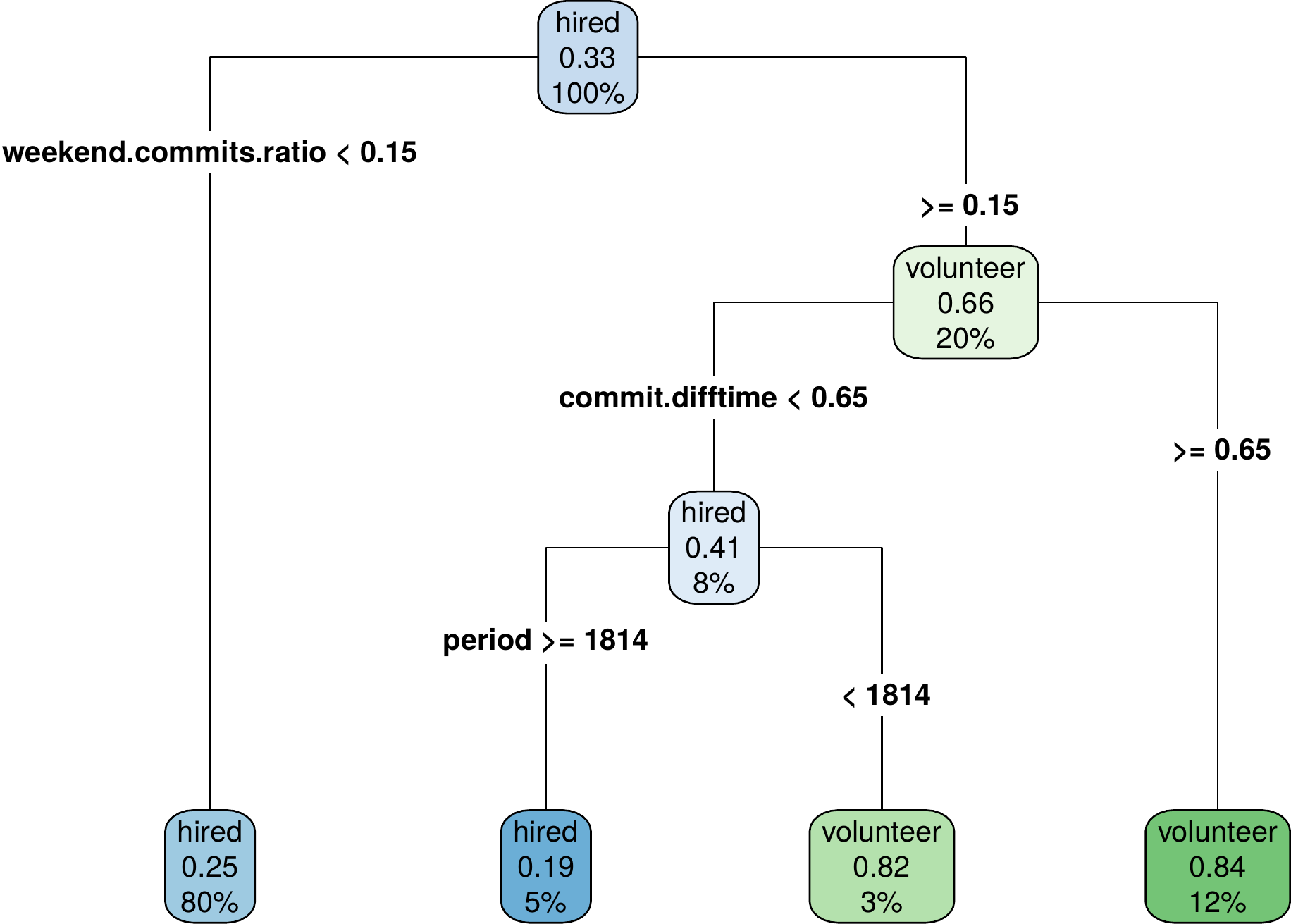}
  \caption{\emph{rpart} decision tree built using all metrics. Each
    node of the tree contains the ratio of volunteer developers and
    the percentage of developers considered.}
  \label{fig:rpart}
\end{figure}

First, when building our models to predict developer employment status
with a repeated k-fold cross validation, we find that
\emph{randomforest} performs better than the two other techniques,
particularly \emph{rpart}. On the other hand, contrarily to
\emph{randomforest}, both \emph{logit} and \emph{rpart} models allow
us to have some insights about the metrics that better predict
employment status. \fig{fig:rpart} depicts the decision tree built by
\emph{rpart} using all metrics as features.

We observe that the first splitting node of the decision tree produced
by \emph{rpart} uses the share of commits made during the weekend. To
further refine the model, it uses the median \emph{time difference}
between successive commits and the length of the \emph{period} of
activity.

Similarly, the most significant coefficient of the \emph{logit} model
is the share of weekend commits (5.94). However the \emph{time
  difference} between commits and the length of the \emph{period} of
activity have much less influence (respectively 0.36 and -0.0004). The
other most important metrics in the \emph{logit} model are the share
of \emph{afternoon} (-1.99), \emph{morning} (-1.08),
\emph{office} (-0.67) and \emph{night} (0.57) activity.

This shows that determining whether a developer is paid, can be done
by relying mostly on the usual commit time of the day or the week and
gives better results than with the simple automatic techniques used in
the literature. Although using email addresses offers a slightly
better precision than any of our methods, there are still a large
number of volunteer developers who uses an e-mail address provided by
the Mozilla foundation. Moreover, this technique has a recall two
times worse than with all of our machine learning models as 30\% of
the hired developers have never used a Mozilla email address.

Using a threshold of 5\% or 95\% of activity during office hours don't
work in the case of Mozilla because all contributors, paid or not,
committed during office hours between 10.4\% and 94.7\% of the
time. In the best case, using a threshold of 5\% is equivalent to
considering that all developers are paid by Mozilla. In the end, using
a threshold of 50\% gives an AUC (0.63) close to the one achieved with
email addresses (0.64) but still far from the AUC of 0.75 obtained
with \emph{randomforest}.

Finally, when trying to predict paid commits instead of paid
developers, while the AUC of email addresses stay unchanged, relying
on regular office hours gives the worst results with an AUC below 0.5
(random guessing) because Mozilla's paid developers work often outside
regular office hours.

For paid commits, we observe that not only
\emph{logit} is now performing worse than the two other models. On the
other hand, \emph{randomforest} is now performing worse (AUC 0.69)
than \emph{rpart}  (AUC 0.72).

%%% Local Variables:
%%% mode: latex
%%% TeX-master: "paper"
%%% End:

%!TEX root = paper.tex

\section{Threats to Validity}\label{sec:threats}

In order to identify developers hired by Mozilla, we manually looked
for information online. Although it allows us to identify a large
amount of the most active developers as hired by Mozilla, we might
have missed developers who do not share online their CV.

We merged developers' identities using a very basic identity merging
technique. We manually checked for false positives in order to avoid
merging the work pattern of two developers as a single one and thus
overestimating their amount of activity. However, there might be false
negatives remaining, which could be particularly problematic in the
case of developers using their work e-mail during office hours and
personal emails outside office hours.

Our study only includes open source projects from Mozilla. The results
obtained are specific to the organization's culture and developer
habits (external validity). Thus we cannot guarantee that our data set
would be representative of the entire open source industry.

%!TEX root = paper.tex

\section{Conclusion and future work}\label{sec:conclusion}

In this paper, we have taken an initial step towards
semi-automatically recognizing paid development in open source
projects. It appears that the best predictors are weekend and evening
work but also time difference between commits. While our models beat
simple automatic classifiers, these results are only an initial step
as the current techniques still can't compete with manually gathered
data.

In the future, we plan to include email-address and commit message
content analysis with natural language processing to improve our
predictions. We will also add other data sources, such as issue
tracker, to improve the amount of information about each
developer. Furthermore, we think giving our machine learning more
information through automated web-scraping as it is a promising way to
further enhance our ability to detect paid development in open source
project and provide more precise results for future MSR studies.

Finally, the main limitation of our study is the lack of
generalizability. Therefore, we want to extend the current study to a
different corpus of open source project. In particular we are
interested in knowing how accurate the technique is in the context of
open source projects where more than one company is involved.

%%% Local Variables:
%%% mode: latexmk
%%% TeX-master: "paper.tex"
%%% End:

%!TEX root = paper.tex

\section*{Acknowledgments}

The authors have been supported by Academy of
Finland grant 298020. 

%!TEX root = debian-era.tex




\bibliographystyle{IEEEtran}
\bibliography{biblio}

\end{document}